\documentclass[10pt,letterpaper]{article}
\usepackage{opex3}
\usepackage{amsfonts}
\usepackage{amssymb}
\usepackage{graphicx}
\usepackage{threeparttable}

\begin{document}

\title{Room temperature photon number resolving detector at telecom wavelengths}

\vskip4pc

\tableofcontents
\clearpage

\title{Room temperature photon number resolving detector at telecom wavelengths}

\author{Enrico Pomarico, Bruno Sanguinetti, Rob Thew, Hugo Zbinden.}

\address{Group of Applied Physics, University of Geneva, 1211 Geneva,
Switzerland.}

\email{enrico.pomarico@unige.ch}

\begin{abstract} Large dynamic
range room temperature photon number resolving (PNR) detectors can be very useful for measuring very low light intensities and for analyzing multiphoton quantum states.
In this paper we present a PNR detector
based on the up-conversion (UC) of telecom signal into visible
wavelength and on its detection by a thermoelectrically cooled multi-pixel silicon avalanche
photodiodode (APD), also known as Silicon Photon Multiplier (SiPM). An efficiency of 4\% is attained and the poissonian statistics of input
coherent states is maintained up to approximately 20 simultaneous detections. The cross-talk effects on the detected signal are estimated in order to properly calibrate the detector. This scheme can be used at arbitrary wavelengths above the visible spectral window with appropriate up-conversion.
\end{abstract}


\bibliography{enricoBn}
\bibliographystyle{unsrt}

\section{Introduction}
Estimating the number of photons in an optical pulse is desirable for the implementation and/or the optimization of a variety of applications.
Several clinical or research activities deal with the analysis of fluorescent compounds for the identification and quantification of chemical reagents or biological molecules. For this task the measurement of the intensity of non repetitive optical pulses over a wide range of values can be improved by the use of photon number resolving detectors (PNR).
Some technological and commercial applications, like optical time domain reflectometry (OTDR), can also gain in
speed and sensitivity if fast and reliable photon counting is used. In the field of quantum information science, photon number resolution can improve the performance of quantum repeater \cite{Simon07} and of QKD
protocols \cite{Felix01,Walton01}, as well as allow linear optics quantum computation \cite{Knill01}.
PNR detectors can also help in the investigation of the properties of multi-photon quantum
states, by means of loss independent measurements of high order correlation functions \cite{Avenhaus10} or with threshold detection conditions \cite{Sekatski09}.

\begin{table}[t]
\centering
\begin{threeparttable}
\begin{tabular}{lcccccc}
\hline \hline
& \small{CIPD}  & \small{TES} &  \small{PND}  & Lin APDs & \small{Non Sat APDs} & \small{This work}  \\

& \cite{Fujiwara06,Fujiwara07} & \cite{Fukuda08,Fukuda09}  & \cite{Divochiy08,Hu09} & \cite{linearAPD09}   &  \cite{Kardynal08}    &                          \\
\hline
\small{Room Temperature}         &   &   &    &  \checkmark  &  \checkmark & \checkmark     \\
\small{High efficiency}          & \checkmark  & \checkmark  &    &    &   &   \\
\small{Fast repetition rates}                 &   &         & \checkmark  & \checkmark  & \checkmark   &  \checkmark       \\
\small{Large dynamic range}     &      &     &     & &       &  \checkmark       \\
\hline
\end{tabular}
\end{threeparttable}

\caption{\small Qualitative comparison between the main PNR detection
approaches in the telecom regime. CIPD = charge integration photodiode; TES = transistor edge sensor; PND = parallel nanowire detector; Lin APDs = linear APDs; Non Sat APDs = Non saturated mode APDs. For large dynamic range we mean a range between 0 and thousands of input photons.}

\label{tab:comparison_PNR}
\end{table}

PNR detectors would be very useful for these applications, if they could estimate photon numbers in single shot measurements. Indeed, the photon number distributions can be reconstructed with just
one single photon detector if a sample of identical photon pulses is available \cite{Rehacek03,Zambra05}.
Actually, even in this case, PNR detectors can make the applications faster, since they provide more
information per pulse with respect to single photon detectors.

Depending on the specific task to be performed, different features are demanded for a PNR detector.
Quantum information tasks require a very efficient and high photon number resolution, but not necessarily a large dynamic range. On the contrary, PNR detectors with limited efficiency can be used for biological or chemical applications and for the investigation of multiphoton
quantum states, but a large dynamic range is highly desirable in this case.

Table \ref{tab:comparison_PNR} qualitatively compares the PNR detectors working at telecom
wavelengths. The charge integration photodiodes (CIPD)
\cite{Fujiwara06,Fujiwara07} and transistor edge sensors (TES)
\cite{Rosenberg05,Fukuda08,Fukuda09} are very efficient and have good photon number resolution. However, they
need cryogenic apparata for cooling, so they are
expensive and cannot work as plug-and-play systems. Moreover, they work at quite slow
repetition rates. Superconducting
parallel nanowires detectors \cite{Divochiy08,Hu09} provide faster responses, but require cryogenic apparata as well. Avalanche photodiodes (APDs) in a
linear mode, even with a very low noise equivalent power (NEP) \cite{linearAPD09},  have a high minimum detectable number of photons (of the order of 10$^{3}$), and
APDs in non-saturated mode
\cite{Kardynal08,Wu09}, where
weak avalanches are measured at an early stage of multiplication,
have a limited efficiency and photon number resolution.

In this paper, we present a PNR detector that works at telecom wavelengths, at room temperature, with a high readout frequency and large dynamic range. This provides a practical detector well suited to measuring low
intensities of light in a single shot fashion or studying
multiphoton quantum states. It is based on
the up-conversion (UC) of telecom photons into the visible regime and their detection by a multi-pixel APD
detector, also known as Silicon Photon Multiplier (SiPM). The use of the UC as an interface for detecting a telecom signal at the visible wavelength is already known \cite{Thew08}, but
this time it is used to exploit the photon counting
capability of the SiPM \cite{Eraerds07,Akiba09}.
Moreover, the UC process allows this approach to be adopted for a wide range of longer wavelengths \cite{Temporao06}.
In the following sections, we describe the detection scheme and data acquisition process.
A characterization of the detector is carried out by measuring its efficiency and noise, before we discuss its photon number counting capability.

\section{The up-conversion multi-pixel APD detector}

\subsection{Scheme of the detector}

\begin{figure}
\begin{center}
\includegraphics[width=0.8\textwidth]{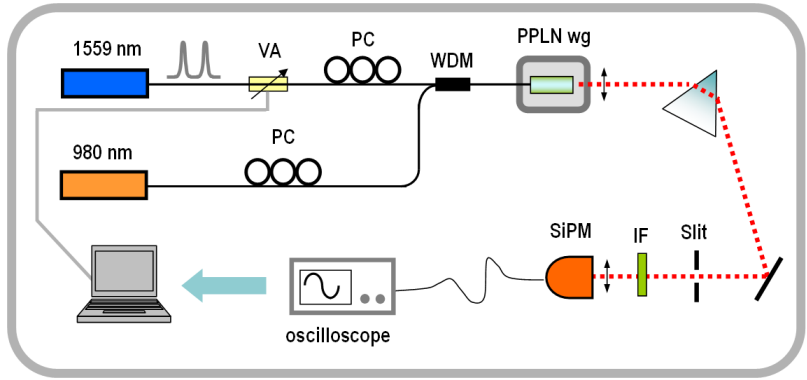}
\caption{\small Schematic of the up-conversion multi-pixel APD
detector. Pulses from a diode laser at 1559\,nm are attenuated (variable attenuator (VA) in the Figure) and
injected, toghether with a pump laser at 980\,nm, into a PPLN waveguide (PPLN wg), where the up-conversion takes place.
Polarization controllers (PC) are used for optimizing the nonlinear process. Light at 600\,nm is then filtered by a dispersion
prism and an interference filter (IF), and detected by the SiPM. The electical signal is registered on the oscilloscope.}\label{fig:setupUP+SiPM}
\end{center}
\end{figure}

A diode laser at 1559\,nm provides pulses of 1\,ns-width that can
be attenuated down to the single photon level, as shown in Figure \ref{fig:setupUP+SiPM}. The telecom signal is
mixed with a cw pump diode laser at 980\,nm (JDS Uniphase) in a
fiber wavelength division multiplexer (WDM). The maximum emitted pump
power is approximately 300\,mW. After the WDM, the telecom signal and the pump are injected into a
periodically poled lithium niobate (PPLN) waveguide (HC
Photonics), where the non linear UC of the signal takes place. The polarization of the input beams is controlled to satisfy the quasi-phase matching (QPM) conditions.
The waveguide has a length of 2.2\,cm, a poling period of
9\,$\mu$m, a nominal normalized internal efficiency of
500\%\,W$^{-1}$\,cm$^{2}$ and its input is fiber pigtailed.

The condition for QPM is obtained at 76.6\,$^{\circ}$C and converts the signal at 1559\,nm to 600\,nm. The upconverted light is collimated and then filtered by a
dispersion prism and an interference filter (IF) at (600$\pm$20)\,nm.
In this way we remove the remaining pump photons and their second
harmonic generation signal at 490\,nm, at which the SiPM is quite efficient. The signal is
then focused onto a free space SiPM (Hamamatsu Photonics
S11028-100(X1)) with 100 APDs arranged on an active area of 1\,mm$^{2}$
with a fill factor of 78.5\%. The choice of the number of pixels is
a compromise between the dynamic range, the efficiency and the
photon number resolution of the detector. Indeed, a SiPM with more than 100
pixels would allow a wider dynamic range, but suffer from a lower photon
sensitivity and efficiency. An internal Peltier cooler allows the
detector to operate down to -$34^{\circ}$C, corresponding to a breakdown voltage of 66.8\,V. The
electrical output is amplified with a 10\,dB amplifier
(Mini-Circuits, 0.1-500\,MHz) and processed by a low-pass filter.

\subsection{Data acquisition}
\begin{figure}
\begin{center}
\includegraphics[width=0.7\textwidth]{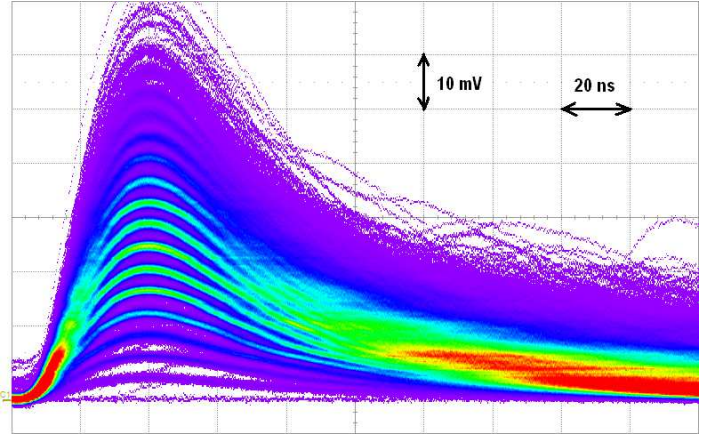}
\caption{\small Superposition of the waveforms of the detected signal on the oscilloscope.}\label{fig:screenshot_oscilloscope}
\end{center}
\end{figure}
\begin{figure}
\begin{center}
\includegraphics[width=1\textwidth]{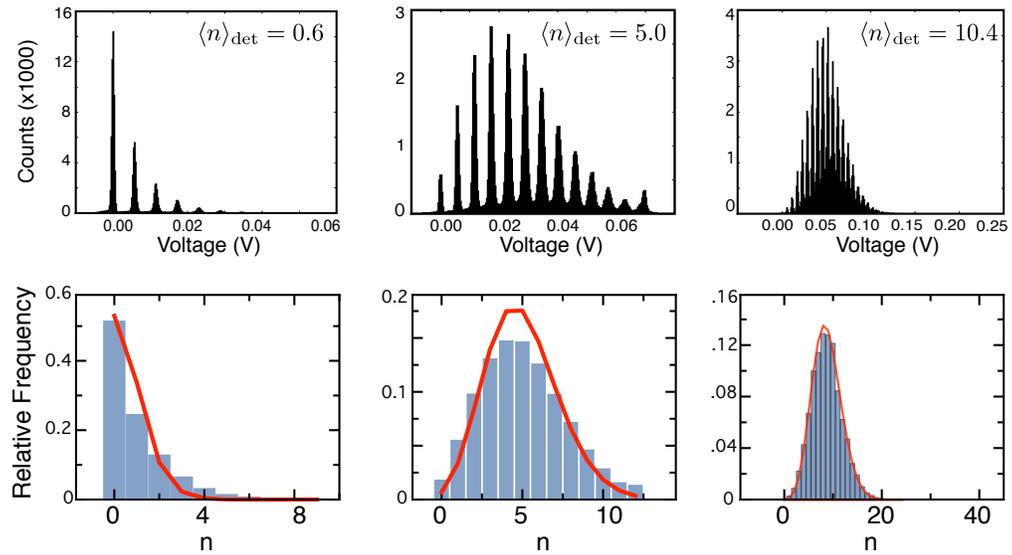}
\caption{\small Top: pulse height histograms for coherent states with mean number simultaneous detections $\langle n\rangle_{det}$ of 0.6, 5.0 and 10.4 respectively.
Bottom: corresponding histograms of the relative frequencies of the simultaneous detections,
fitted by poissonian distributions
(red lines).
$n$ is the number of simultaneous detections. }\label{fig:signal&histogram}
\end{center}
\end{figure}

Telecom wavelength optical pulses are sent to the detector with a 100\,kHz
repetition rate (see Figure
\ref{fig:setupUP+SiPM}). Their intensity can be set by varying the
attenuation automatically by software. The signal detected by the SiPM is measured with a 500\,MHz oscilloscope, triggered by a PIN photodiode that detects a
portion of light from the optical source. The amplitude of the SiPM
output, in a relatively low photon number regime, is approximately proportional to the number of simultaneous
avalanches in the detector \cite{Eraerds07,Akiba09}. In Figure \ref{fig:screenshot_oscilloscope} the superposition of
the waveforms of the detected signal on the oscilloscope is shown. Their amplitudes take
quite distinct values that are distributed according to the poissonian
distribution of the number of detected photons. Their duration is approximately 50\,ns.

For each coherent state sent to the detector, we register 10$^{5}$ traces from the oscilloscope and measure the
height of the signal for each trace. The
upper part of Figure \ref{fig:signal&histogram} shows the pulse height
histograms for coherent states with mean number of simultaneous detections $\langle n \rangle_{det}$ of 0.6, 5.0 and 10.4 respectively.
For the first two states one can notice a very good separation between the different amplitude levels.
This confirms the good sensitivity of the SiPM and the fact that all the pixels have similar gains. The distinguishability of the
peaks gets worse for an increasing number of simultaneous
detections, because the uncertainty on the output voltage produced by the firing of $N$
pixels is approximately proportional to $\sqrt{N}$ times the standard deviation of the signal associated to one detection. For the coherent state with $\langle n\rangle_{det}=$10.4, the peaks are still quite distinct, despite
the appearance of a gaussian background due to the overlap of the
tails of the detection signals.

For each state the data are organized in a histogram of
the relative frequencies of the simultaneous detections, from which we determine the mean ($\langle n \rangle_{det}$) and the variance ($\sigma^2_{det}$) values. The data are then fitted with a poissonian
distribution, as shown in the three graphs in the lower part of the
Figure \ref{fig:signal&histogram}.


\subsection{Efficiency and noise characterization}

The detector efficiency depends on the efficiency of the whole UC and filtering process and of the SiPM detection efficiency.
With the maximum available pump power, the signal at
600\,nm measured after the IF is only the 11\% of the telecom
light sent to the input of the detector, due to the the losses of the system. Only approximately 23\% of the telecom light is coupled into the waveguide. A nominal value of 65\% should be found for the coupling efficiency at telecom wavelengths. This discrepancy is due to the deterioration in time of the input fiber pigtailing.  The number of telecom photons can be upconverted in the PPLN waveguides in principle with a 100\% efficiency with a greater pump power \cite{Roussev04}. However, the pump power effectively available for the UC is limited by the losses due to fibre splices, to the WDM and also to the waveguide coupling. Finally, the IF has a
transmission of 85\%.

To analyse the detection system, we send telecom wavelength coherent pulses to the detector with different mean
photon numbers $\langle n \rangle_{in}$ at 100\,kHz. At this frequency the effects of afterpulsing are negligible, but
in principle repetition rates of up to 20\,MHz are possible. We measure the efficiency of the SiPM and of
the total detector as
$\eta=\frac{\langle n \rangle_{det}}{\langle n \rangle_{in}}$, where $\langle n \rangle_{det}$ and $\langle n \rangle_{in}$ are the mean numbers of simultaneous detections and of input photons per pulse, respectively. We measured for the SiPM an efficiency value of $\eta_{SiPM}=$36\% with the bias voltage at 68.1\,V (corresponding to an excess bias voltage of 1.3\,V). The excess bias voltage
is calculated as the difference between the bias and the breakdown voltage. Notice that the definition of $\eta$ includes effects of cross-talk, that we will explain in detail in the next paragraph. By counting the total number of detection events, regardless of the amplitude of the avalanches, we can estimate the quantum efficiency of the SiPM to be 24\%.
The detector is set at a temperature of -34\,$^{\circ}$C to
minimize the thermal noise.

\begin{figure}
\begin{center}
\includegraphics[width=0.6\textwidth]{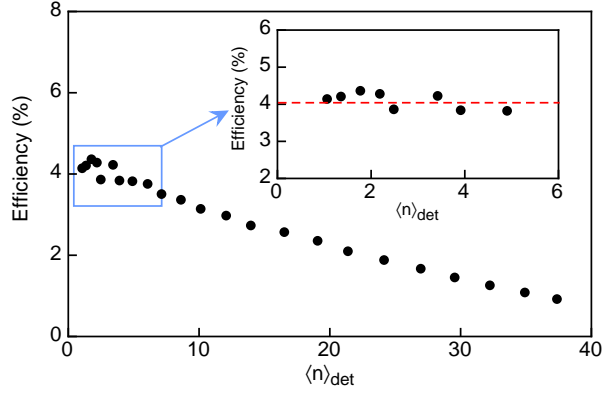}
\caption{\small Efficiency $\eta_{tot}$ of the UC multi-pixel APD detector as a
function of the mean numbers of simultaneous detections $\langle n \rangle_{det}$.
}\label{fig:efficiency}
\end{center}
\end{figure}

\begin{figure}
\begin{center}
\includegraphics[width=0.6\textwidth]{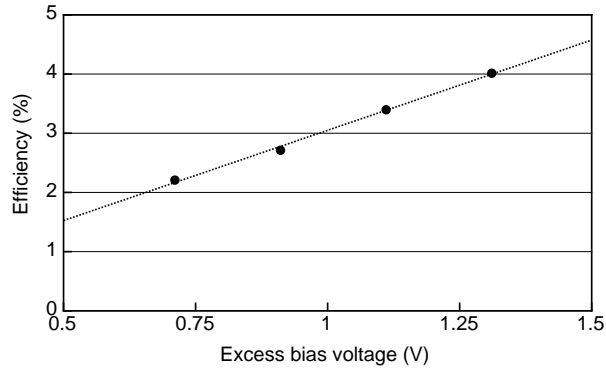}
\caption{\small Efficiency $\eta_{tot}$ of the UC multi-pixel APD
detector as a function of the excess bias voltage applied to the SiPM. The dotted line corresponds to a linear fit of the data.
}\label{fig:efficiency_noiseVSbias}
\end{center}
\end{figure}

Figure \ref{fig:efficiency} plots the total efficiency $\eta_{tot}$ of the total detector as a function of $\langle n \rangle_{det}$,
for an excess bias voltage of 1.3\,V. The
detected signal grows linearly with the input signal up to approximately 6 simultaneous detections,  and the efficiency of the detector is approximately constant at 4\%,
which is what we expect from the individual
values for the UC and SiPM efficiency. For higher values of $\langle n\rangle _{det}$, $\eta_{tot}$ decreases
because of the saturation of the SiPM. Figure \ref{fig:efficiency_noiseVSbias} shows the linear trend of
the detector efficiency (for low numbers of simultaneous detections) as a function of the excess bias voltage applied to the SiPM.
The efficiency shows an increase of (0.31$\pm$0.01)\% per 0.1\,V of excess bias voltage.
The amplitude of the avalanches also increases in a linear way with (0.39$\pm$0.01)\% per 0.1\,V of excess bias voltage.

The noise of the detector has two different origins: the electronic noise of the SiPM, which is not produced by incident photons, and the detection of photons at 600\,nm
originating from other non linear processes (discussed in
\cite{Thew06}), which take place inside the waveguide.
We calculate the histogram of the detection frequency in the absence of the input telecom signal, from which we find that the noise corresponds to $\langle n\rangle_{det}$=0.023. This value can be interpreted as a noise probability per shot. In order to evaluate the minimal sensitivity of our detector, we divide it by the efficiency of the detector and obtain 0.58. This is the approximate number of incident photons that give the same mean number of detections, if the detector had no noise and if they were sent in pulses of 1\,ns width.
The majority of this noise is due to the
UC. Indeed, only 2\% of the total noise is due to the SiPM.

\subsection{Multi-photon detection}

The multi-photon counting capability of a PNR detector could be in principle characterized by performing a quantum tomography of the detector, according to the method described in \cite{Feito09}. However, in the case of large dynamic range PNR detectors, the number of parameters to find numerically is very large, because probe states with large mean photon numbers have to be used for the characterization of the detector. This makes the numerical problem extremely complex and unstable. Therefore, more practical approaches need to be found for the characterization of PNR detectors in a large photon number regime. In our case we investigate the preservation in the detected signal of the statistical properties of the input coherent states.
If one sends Fock states to a
perfectly efficient PNR detector, with a sufficiently high photon number resolution, distinct outcomes for
each state should be expected. In the case of a single shot
measurement, one obtains a specific output for only one specific
Fock state. For a coherent state a poissonian distribution of outcomes is obtained, thus one can infer the number of
photons of the initial pulse only up to the intrinsic uncertainty
given by the poissonian statistics.

\begin{figure}
\begin{center}
\includegraphics[width=0.8\textwidth]{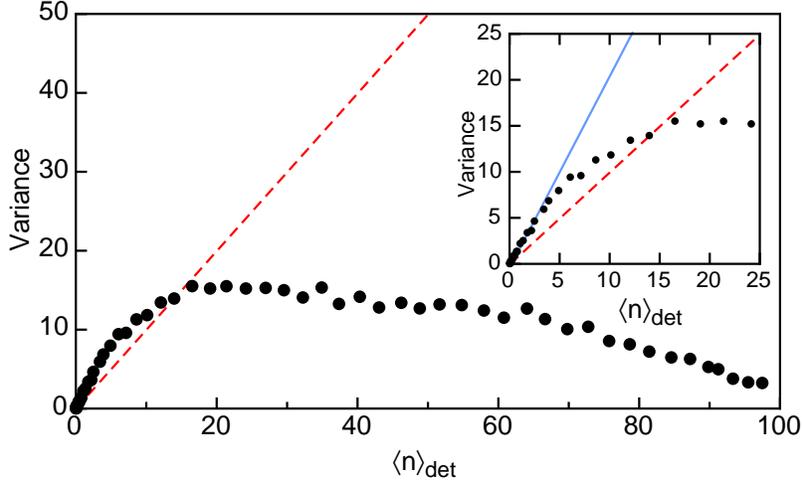}
\caption{\small Experimental variances $\sigma^2_{det}$ as a
function of the corresponding mean numbers of simultaneous detections $\langle n\rangle_{det}$.
The dashed red line corresponds to ideally detected coherent states, for which $\sigma^2_{det}=\langle n\rangle_{det}$.
In the inset the data for low $\langle n\rangle_{det}$ are shown. The blue line corresponds to the fit of the first data for the
calculation of the cross-talk probability of the SiPM.}\label{fig:varianceVSmean}
\end{center}
\end{figure}

A good PNR
detector should preserve the statistics of the input states: the equality between the mean number of simultaneous detections
$\langle n\rangle_{det}$ and the variance of the experimental data distributions
$\sigma^2_{det}$ should be verified. In Figure \ref{fig:varianceVSmean} we plot the
experimental variances $\sigma^2_{det}$ as a function of the
corresponding mean numbers of simultaneous detections $\langle n\rangle_{det}$.

The dashed red line in Figure \ref{fig:varianceVSmean} corresponds to the condition of perfect poissonian detection distribution, that is $\langle n\rangle_{det}=\sigma^2_{det}$. The variances of the experimental
distributions decrease for more
than 20 simultaneous detections due to the saturation of
the SiPM. The experimental
distributions of the number of simultaneous detections becomes narrower as $\langle n\rangle_{det}$
approaches 100, the number of detector pixels.

The deviation of the experimental variances from the condition of perfectly detected coherent states, for $\langle n \rangle_{det}$ less than 15 (inset of the Figure \ref{fig:varianceVSmean}), deserves further discussion.
We attribute this fact mainly to optical cross-talk effects
between adjacent pixels. The avalanche process in one APD can induce the firing of one ore more neighbouring pixels,
as the carriers released during the avalanche can emit radiative photons (the emission probability is approximately one
photon per 10$^{5}$ carriers) \cite{Rech08}. We determine the cross-talk probability in our detector by considering the recent analytic results
reported in \cite{Vinogradov09}. The authors of \cite{Vinogradov09} assume that the pixels can fire because of a primary origin,
the detection of a photon, or a secondary one, for cross-talk or afterpulsing, which is essentially random (in our case the afterpulses are negligible.).
Because of the latter effect, the experimental distributions of the number of simultaneous detections corresponding to input coherent states are compound poissonians with mean values $\langle n \rangle_{det} =\frac{\langle n \rangle}{1-p}$ and variances $\sigma^2_{det}=\frac{\langle n \rangle (1+p)}{(1-p)^2}$, where $p$ is the probability of firing for secondary events and $\langle n \rangle$ is the mean of the distribution in the absence of cross-talk. According to this model, the mean and the variance of the compound distributions increase in the presence of the secondary effect. The Fano factor $F$, which is the ratio
between the variance of the distribution of the simultaneous detections $\sigma_{det}^2$ and the mean number of this distribution $\langle n\rangle_{det}$, is thus expressed by the relation $F=\frac{1+p}{1-p}$, from which $p$ can be estimated. We measure $F$ from the slope of the line fitting the
data in the inset of Figure \ref{fig:varianceVSmean}. We limit the fit to low numbers of $\langle n \rangle_{det}$, where the detector is unaffected by saturation. We obtained a cross-talk probability of (31.4 $\pm$ 0.6)\% at an excess bias voltage of 1.3\,V.

\begin{figure}
\begin{center}
\includegraphics[width=0.6\textwidth]{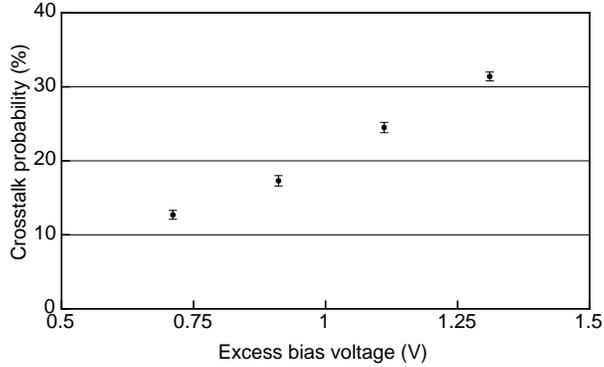}
\caption{\small Cross-talk probability as a function of the excess bias voltage applied to the SiPM. The error bars represent the statistical errors.}\label{fig:crosstalkVSbias}
\end{center}
\end{figure}

The cross-talk probability increases as a function of the excess
bias voltage (Figure \ref{fig:crosstalkVSbias}).
A similar effect is reported in \cite{Mizumura09}. The probability for a pixel
to fire due to cross-talk is related to its
efficiency, but also on the quantity of photons released during the avalanche of a neighbouring pixel, which depends on the amplitude
of the avalanches. Both factors increase with the excess bias voltage.

The cross-talk probability affects the values of efficiency $\eta_{SiPM}$ and $\eta_{tot}$. The relation between the value of $\eta$ and the quantum efficiency of the detector $QE$ can be expressed in our case as $\eta=QE(1+p')$, where $p'$ is the total probability of having detections due to cross-talk of second and multiple order, that is $p'=\sum_{n=1}^{\infty} p^n=\frac{p}{1-p}$, which in our case is (45.7 $\pm$ 1.3)\% at an excess bias voltage of 1.3\,V for the value of $p$ determined before.
Therefore, we obtain a value of $QE_{SiPM}$=25\% for the SiPM and of $QE_{tot}$=2.7\% for the total detector at an excess bias voltage of 1.3\,V. Considering the systematic errors in the estimation of the input photon number, the value of $QE_{SiPM}$ is in good agreement with what we measured directly. Correcting the effects of cross-talk in the detected signal and in the efficiency enables one to perform the calibration of the detector. In other terms it allows one to properly estimate the number of photons in an optical pulse from the detected signal.

\section{Conclusion}
In this paper we presented a telecom wavelength PNR detector with a large
dynamic range, working at room temperature.
It is based on the UC of a telecom signal into the visible wavelength regime
and detection by a thermoelectrically cooled SiPM. This kind of detector can be used for a wide range of infrared wavelengths with appropriate up-conversion.
The efficiency is limited to 4\% due to the deteriorated waveguide coupling, however, optimized systems have achieved pigtailing losses $<0.5$dB with further improvements possible \cite{Roussev04}. The detector preserves the poissonian statistics of the input coherent states up to 20 simultaneous detections, corresponding to approximately 740 input photons if one considers the quantum efficiency of the detector. The effects of cross-talk in the detected signal can be theoretically estimated. Their correction allows the proper calibration of the detector.

\section*{Acknowledgments}
This work is supported by the Swiss NCCR "Quantum Photonics". We would like to thank Olivier
Guinnard and Claudio Barreiro for their technical support and Patrick Eraerds and Jun Zhang for their
valuable comments and suggestions.

\end{document}